\documentclass[aps,twocolumn]{revtex4}%
\usepackage{amsfonts}
\usepackage{amsmath}
\usepackage{amssymb}
\usepackage{graphicx}%
\begin{document}
\title{Would quantum entanglement be increased by anti-Unruh effect?}
\author{Taotao Li}
\author{Baocheng Zhang}
\email{zhangbc.zhang@yahoo.com}
\affiliation{School of Mathematics and Physics, China University of Geosciences, Wuhan
430074, China}
\author{Li You}
\affiliation{State Key Laboratory of Low Dimensional Quantum Physics, Department of
Physics, Tsinghua University, Beijing 100084, China}

\begin{abstract}
We study the \textquotedblleft anti-Unruh effect\textquotedblright\ for an
entangled quantum state in reference to the counterintuitive cooling
previously pointed out for an accelerated detector coupled to the vacuum. We
show that quantum entanglement for an initially entangled (spacelike
separated) bipartite state can be increased when either a detector attached to
one particle is accelerated or both detectors attached to the two particles
are in simultaneous accelerations. However, if the two particles (e.g.,
detectors for the bipartite system) are not initially entangled, entanglement
cannot be created by the anti-Unruh effect. Thus, within certain parameter
regime, this work shows that the anti-Unruh effect can be viewed as an
amplification mechanism for quantum entanglement.

\end{abstract}
\maketitle

\section{Introduction}

It is widely accepted by now that an observer with uniform acceleration $a$ in
the Minkowski vacuum of a free quantum field would feel a thermal bath of
particles at the temperature $T={\hbar a}/{(2\pi ck_{B})}$ \cite{wgu76}. This
effect, discovered in 1976 by Unruh, implicates that the particle content of a
quantum field is observer dependent. It has been since digested and extended
to many different situations (see the review \cite{chm08} and references
therein). In one famous application to the Unruh-DeWitt detector \cite{bsd79},
it is found that a quantum system consisting of a detector uniformly
accelerating in Minkowski vacuum can sense the thermal emission and thus cause
decoherence due to the coupling with the thermal field.

However, it is found recently that a particle detector in uniform acceleration
coupled to the vacuum can cool down with increasing acceleration \textit{under
certain conditions}. This scenario is opposite to that gives the celebrated
Unruh effect, and has been appropriately named the anti-Unruh effect
\cite{bmm16}. This initial discussion was based on a point-like two-level
system and the transition probability was found to decrease with increasing
acceleration rather than the expected nominal increasing dependence. Although
the initial calculation is made in Ref. \cite{bmm16} for accelerated detectors
coupled to a massless scalar field either in a periodic cavity or under a
hard-IR momentum cutoff for the continuum, it is also showed to represent a
general stationary mechanism that remains stable under the disturbance of
additional conditions, instead of being a sheer transient phenomenon
\cite{gmr16}. Thus, like the Unruh effect, the anti-Unruh effect constitutes a
significant breakthrough in our understanding.

Since the anti-Unruh effect can exist under a stationary state satisfying
Kubo-Martin-Schwinger (KMS) condition \cite{rk57,ms59,fjl16} and is
independent on any kind of boundary conditions \cite{bmm16,gmr16}, what is its
difference from Unruh effect? Although the physically essential reasons remain
to be explored, some important elements, like the interaction time, the
detector's energy gap, the mass of the quantum field, etc, had been discussed
carefully to distinguish the two situations in the earlier works
\cite{bmm16,gmr16}. For example, the anti-Unruh effect could appear when the
interaction timescale is far away from the timescale associated to the
reciprocal of the detector's energy gap. It is also noted that these
discussions were made under the background of Unruh-DeWitt detector, so a
necessary step is to check whether the anti-Unruh effect can also be applied
under some other situations, i.e. whether it has any influence on quantum
entanglement, and whether the influence is the same as or different from that
of the Unruh effect.

The influence of the Unruh effect on quantum entanglement has been subjects of
many studies \cite{chm08}. A recent study finds that a maximally entangled
quantum state in an inertial frame becomes less entangled to an observer in
relative acceleration \cite{fm05}. This degradation of entanglement as well as
the possibility of its sudden death has also been investigated for spacelike
separated observers with the same acceleration \cite{dss15}. The results from
these studies have been further extended to different situations
\cite{amt06,ml09,mgl10,wj11,ses12,bfl12,ro15}, and they help to establish the
general conclusion that entanglement is also observer dependent. It decreases
for accelerating observers, or accelerating observers only have partial access
to the information encoded in the quantum entanglement. However, the real
reason why the inertial observers would measure more (the maximal amount of)
entanglement than others remains to be fully understood, since the notion of
acceleration is always relative. In particular, the possibility of enhanced
entanglement for accelerating observers is not ruled out. Indeed, several
studies have concluded that the Unruh effect can actually lead to enhanced
quantum entanglement by coupling one or two detectors into the local quantum
fields even if they were spacelike separated \cite{rrs05,sm09,mm11}. Since no
local quantum operations can increase the amount of entanglement between two
parties of a quantum system \cite{nc00}, this entanglement enhancement is
speculated to be extracted from the quantum entanglement of vacuum with which
the accelerated detectors interacted, by a mechanism similar to entanglement
swapping \cite{zze93,pbz98}. \ In particular, these phenomena of enhancement
of entanglement didn't represent a stationary mechanism.

When we attempt to discuss the influence of anti-Unruh effect on quantum
entanglement between two detectors, it is expected that the phenomena of the
enhancement of entanglement could appear in a case with the stationary state
but has to avoid the influence of the vacuum entanglement. So in this paper,
we take a product state for the vacuum but the results are still valid for the
general entangled vacuum state. Thus, it establishes our problem for
considering whether quantum entanglement between two spacelike separated
parties can be enhanced or not when the anti-Unruh effect is enforced. The
answer to this question could profoundly change our understanding to the
anti-Unruh effect.

This paper is organized as follows. First, in section II we review the all
important decoherence factors due to the anti-Unruh effect, instead of using
the transition probability in earlier studies. This is followed in section III
by the discussions on entanglement enhancement due to the anti-Unruh effect
for several different situations, where we take a product state for the vacuum
in order to avoid the confusion with the mechanism of entanglement swapping.
Finally we consider the case of an initial product state in section IV, which
is followed by the conclusion in section V.

\section{The anti-Unruh effect}

We first briefly review the anti-Unruh effect presented in Ref. \cite{bmm16}
with the Unruh-DeWitt (UDW) model, but with the decoherence factor instead of
the initial use of the transition probability and with the massive field
instead of the initial massless field. Starting with the consideration that
constitutes a scalar field $\phi$ interacting with a point-like two-level
quantum system, or a qubit (for short). It can be easily generalized to more
complex situations such as a quantum oscillator \cite{bmm13} as confirmed with
KMS conditions for thermal equilibrium \cite{gmr16}. The ground $\left\vert
g\right\rangle $ and excited $\left\vert e\right\rangle $ states of the qubit
are separated by an energy gap $\Omega$ while experiencing accelerated motion
in a vacuum cavity. The interaction Hamiltonian for this ($1+1$)-dimension
model is given by
\begin{equation}
H_{I}=\lambda\chi\left(  \tau/\sigma\right)  \mu\left(  \tau\right)
\phi\left(  x\left(  \tau\right)  \right)  , \label{udwi}%
\end{equation}
with $\lambda$ the coupling strength. $\tau$ is the qubit's proper time along
its trajectory $x\left(  \tau\right)  $, $\mu\left(  \tau\right)  $ is the
qubit's monopole momentum, and $\chi\left(  \tau/\sigma\right)  $ is a
switching function that is used to control the interaction time scale $\sigma
$. $\phi(x(\tau))$ is the scalar field related to the vacuum. For a qubit
accelerating in a vacuum cavity, the evolution of the total quantum state is
determined perturbatively by the unitary operator which up to first order is
given by,
\begin{equation}
U=I-i\int d\tau H\left(  \tau\right)  +O\left(  \lambda^{2}\right)  .
\end{equation}
Within the first-order approximation and in the interaction picture, this
evolution is described by \cite{bmm16}
\begin{align}
U\left\vert g\right\rangle \left\vert 0\right\rangle  &  =C_{0}\left(
\left\vert g\right\rangle \left\vert 0\right\rangle -i\eta_{_{0}}\left\vert
e\right\rangle \left\vert 1_{k}\right\rangle \right)  ,\nonumber\\
U\left\vert e\right\rangle \left\vert 0\right\rangle  &  =C_{1}\left(
\left\vert e\right\rangle \left\vert 0\right\rangle +i\eta_{_{1}}\left\vert
g\right\rangle \left\vert 1_{k}\right\rangle \right)  , \label{foe}%
\end{align}
where $k$ denotes the mode of the ($1+1$)-dimension scalar field with
(bosonic) annihilation (creation) operator $a_{k}$ ($a_{k}^{\dag}$),
$a_{k}\left\vert 0\right\rangle =0$ and $a_{k}^{\dag}\left\vert 0\right\rangle
=\left\vert 1_{k}\right\rangle $. $\eta_{_{0}}=\lambda\int dkI_{+,k}$ and
$\eta_{_{1}}=\lambda\int dkI_{-,k}$ are related to the excitation and
deexcitation probability of the qubit where $I_{\pm,k}$ is given as $I_{\pm
,k}=\int_{-\infty}^{\infty}\chi\left(  \tau/\sigma\right)  \exp[\pm
i\Omega\tau+i\omega t\left(  \tau\right)  -ikx\left(  \tau\right)  ]{d\tau
}/{(\sqrt{4\pi\omega})}$. $t\left(  \tau\right)  =a^{-1}\sinh(a\tau)$ and
$x\left(  \tau\right)  =a^{-1}\left(  \cosh(a\tau)-1\right)  $ is the
trajectory of the accelerating qubit with acceleration $a$. $C_{0,1}%
=1/\sqrt{1+\eta_{_{0,1}}^{2}}$ is the state normalization factor. It is worth
pointing out that the periodic boundary conditions are not used for $I_{\pm
,k}$, thus this review in essence generalizes the case considered in Ref.
\cite{bmm16}. We consider massive field with e.g. $\omega=\sqrt{k^{2}+m^{2}}$
as in Ref. \cite{gmr16} so that the anti-Unruh effect discussed here will not
be constrained by the finite interaction time and its validity can be extended
to situations where the detector is switched on adiabatically over an infinite
long time. Without loss of generality, $m=1$ is used for all numerical calculations.

\begin{figure}[ptb]
\centering
\includegraphics[width=3.35in]{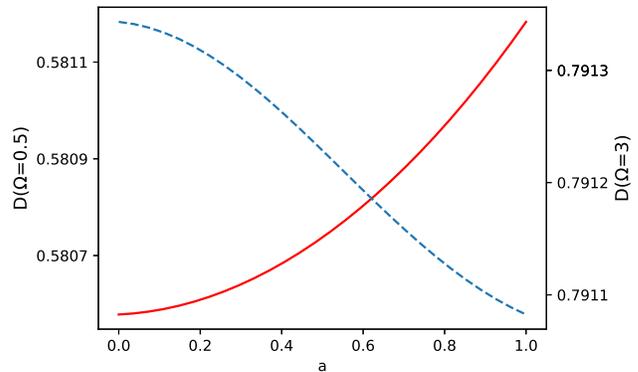} \caption{(Color online) The
dependence of decoherence factor $D$ on the acceleration $a$. The model
parameters employed are $\lambda=0.1$ and $\sigma=0.4$. The red solid line
denotes enhanced coherence with acceleration at $\Omega=0.5$ (referenced to
the left vertical axis), while the blue dashed line with respect to the right
vertical axis is for decreased coherence with acceleration at $\Omega=3$. }%
\label{fig1}%
\end{figure}

\begin{figure}[ptb]
\centering
\includegraphics[width=3.35in]{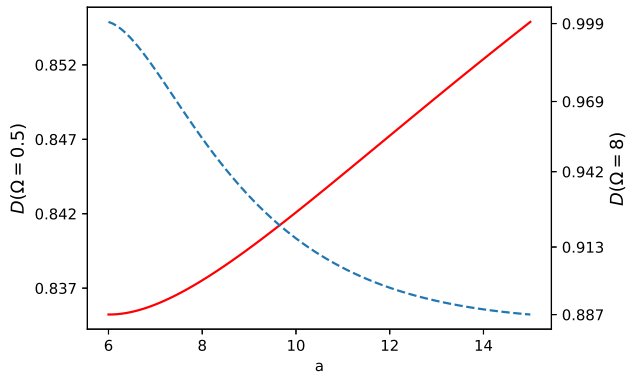} \caption{(Color online) The same as
in Fig. \ref{fig1} but for $\sigma\rightarrow\infty$. The blue dashed line now
refers to $\Omega=8$ for decreased coherence with acceleration. }%
\label{fig2}%
\end{figure}

For an initial qubit state $\left\vert \psi_{i}\right\rangle =\left(
\alpha\left\vert g\right\rangle +\beta\left\vert e\right\rangle \right)
\left\vert 0\right\rangle $ with the complex amplitudes $\alpha$ and $\beta$
satisfying the normalization $\left\vert \alpha\right\vert ^{2}+\left\vert
\beta\right\vert ^{2}=1$, its evolution according to Eq. (\ref{foe}), after
interacting with the scalar field, leads to the state
\begin{equation}
\left\vert \psi_{f}\right\rangle =\alpha\left\vert g\right\rangle \left\vert
\psi_{0}\right\rangle +\beta\left\vert e\right\rangle \left\vert \psi
_{1}\right\rangle ,
\end{equation}
with $\left\vert \psi_{0}\right\rangle =C_{0}\left\vert 0\right\rangle
+i({\beta}/{\alpha})C_{1}\eta_{_{1}}\left\vert 1_{k}\right\rangle $ and
$\left\vert \psi_{1}\right\rangle =C_{1}\left\vert 0\right\rangle -i({\alpha
}/{\beta})C_{0}\eta_{_{0}}\left\vert 1_{k}\right\rangle $. The loss of
coherence for the qubit is measured by the decoherence factor $D=\left\vert
\left\langle \psi_{0}\right\vert \left.  \! \psi_{1}\right\rangle \right\vert
$, which is related to the purity of the reduced qubit density matrix after
evolution \cite{blu13}. $D=0$ means that the qubit state becomes completely
mixed and loses coherence completely, while $D=1$ means that the state remains
pure. For any other value of $D$ in between $\in(0,1)$, the coherence of the
qubit is partially lost. Figure \ref{fig1} shows the dependence of $D$ on $a$
for our model with $\alpha=\beta=\sqrt{2}/{2}$ for the initial state. It
explicitly shows that under suitable conditions, the qubit state coherence is
enhanced, which implicates that accelerated motion can potentially purify a
quantum state. One should of course remain cautious as in this example the
initial coherence is not simply given by $D=1$ for $a=0$, due to the influence
of the switching function \cite{bmm13,ls08}. However, the enhancement cannot
be simply credited to the finite time interaction either. For significantly
prolonged interaction times, this enhancement is observed to stay, however, as
shown by the red solid line in Fig. \ref{fig2}. Although Figs. \ref{fig1} and
\ref{fig2} correspond to different values of $a$ and $\Omega$ because the
switching function stops working when the interaction time becomes infinite,
Fig. \ref{fig2} firmly establishes that the anti-Unruh effect exists even for
an infinite time interaction in this case.

The change to qubit coherence considered above can be extended to a many-body
quantum state, e.g. in a tri-partite state $\left\vert \psi_{3i}\right\rangle
=\left(  \alpha\left\vert g\right\rangle \left\vert g\right\rangle \left\vert
g\right\rangle +\beta\left\vert e\right\rangle \left\vert e\right\rangle
\left\vert e\right\rangle \right)  $ which reduces to the famous GHZ state
\cite{ghz89} for $\alpha=\beta=\sqrt{2}/{2}$. After evolution under the
analogous interaction model, this state becomes $\left\vert \psi
_{3f}\right\rangle =\alpha\left\vert g\right\rangle \left\vert g\right\rangle
\left\vert g\right\rangle \left\vert \psi_{0}\right\rangle +\beta\left\vert
e\right\rangle \left\vert e\right\rangle \left\vert e\right\rangle \left\vert
\psi_{1}\right\rangle $, whose coherence is described by the same decoherence
factor $D$ and exhibits similar $a$-dependence. Furthermore, with two or three
qubits (of the tri-qubit system) in acceleration simultaneously, the changes
to the state coherence show similar behaviors which can be confirmed
straightforwardly albeit after a little bit of more complicated calculations.
In what follows, we will study another one question concerning the change to
quantum entanglement among different qubits when there are qubits coupled to
the quantum vacuum. This is different from the change of the coherence which
is for the whole state. We hope to understand whether this change of
entanglement between two qubits has the similar phenomena to the coherence
discussed here.

\section{Enhancement of entanglement}

The previous section shows that single qubit coherence can be enhanced by the
anti-Unruh effect. We now take one step further and consider the influence of
anti-Unruh effect on quantum entanglement \cite{hhh09}. This study is timely
since entanglement is viewed as a resource for quantum information science. We
will consider two causally separated qubits in an entangled state, where each
qubit is independently accelerating in a vacuum cavity and assumes the same
coupling with the scalar field in its respective (spatial) place by the same
process presented in Eq. (\ref{foe}). Different from the steps reviewed above
for treating the coherence factor for a single qubit, we will employ measures
for quantum entanglement to discuss the influence on qubit entanglement due to
the anti-Unruh effect in the following.

In the general scenario we consider, the initial two-qubit state is a pure one
in flat spacetime. When one of the qubits is accelerating in a vacuum cavity,
the bipartite state becomes mixed and can be measured by the decoherence
factor. Concerning quantum entanglement, it is well understood for any
bipartite pure state or mixed states of two qubits. There exist several
established entanglement measures for mixed states of two qubits, which
include the widely adopted concurrence \cite{ww98}, logarithmic negativity
\cite{vw02,mbp05}, and mutual information \cite{nc00}. Bipartite mutual
information is a measure of total correlation between the two subsystems of a
quantum state, i.e., the sum of its quantum entanglement and classical
correlation. So in this paper, we choose to adopt concurrence and logarithmic
negativity as measures to the change of the bipartite entanglement due to acceleration.

The logarithmic negativity $E_{N}$ is a nice measure for mixed state, although
it fails to reproduce the entropy of entanglement of pure state like most
other entanglement measures. It is defined as
\begin{equation}
E_{N}=\log_{2}\left(  2N+1\right)  , \label{lne}%
\end{equation}
and bounded by $0\leqslant E_{N}\leqslant1$, where the negativity $N$ is
computed according to $N=\sum_{i}\left(  \left\vert \xi_{i}\right\vert
-\xi_{i}\right)  /2$ with $\xi_{i}$ the $i$-th eigenvalue of the partial
transpose of the bipartite state density matrix. Concurrence defined by
\begin{equation}
C\left(  \rho\right)  =\max\{0,\lambda_{1}-\lambda_{2}-\lambda_{3}-\lambda
_{4}\} \label{mui}%
\end{equation}
is also a widely used entanglement measure for bipartite mixed state, where
$\lambda_{1}$, $\lambda_{2}$, $\lambda_{3}$, $\lambda_{4}$ are the eigenvalues
of the Hermitian matrix $\sqrt{\sqrt{\rho}\widetilde{\rho}\sqrt{\rho}}$ with
$\widetilde{\rho}=\left(  \sigma_{y}\otimes\sigma_{y}\right)  \rho^{\ast
}\left(  \sigma_{y}\otimes\sigma_{y}\right)  $ the spin-flipped state of
$\rho$, $\sigma_{y}$ being the y-component Pauli matrix, and the eigenvalues
listed in decreasing order.

We now consider the change to entanglement due to the anti-Unruh effect when
qubits are in acceleration. The initial state is assumed to take the form
\begin{align}
\left\vert \Psi_{i}\right\rangle =\left(  \alpha\left\vert g\right\rangle
_{A}\left\vert e\right\rangle _{B}+\beta\left\vert e\right\rangle
_{A}\left\vert g\right\rangle _{B}\right)  \left\vert 0\right\rangle
_{A}\left\vert 0\right\rangle _{B}, \label{eqs}%
\end{align}
again with the complex coefficients satisfying $\left\vert \alpha\right\vert
^{2}+\left\vert \beta\right\vert ^{2}=1$ and the vacuum as in a product state.
Thus, the entanglement swapping between the vacuum and the entangled state of
the two qubits far away from each other is excluded, and the contribution to
the entanglement change due to the anti-Unruh effect can be singled out to
study entanglement enhancement. In what follows, we will consider two
different cases.

\subsection{$B$-qubit in acceleration}

\begin{figure}[ptb]
\centering
\includegraphics[width=3.65in]{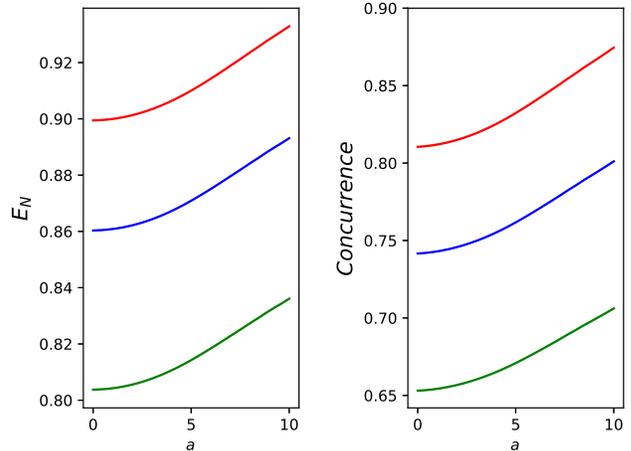} \caption{(Color online) The
negativity (left panel) and the concurrence (right panel) as a function of $a$
when one of the qubits is in acceleration. The parameters correspond to the
cases of enhanced entanglement with $\lambda=0.1$, $\sigma=0.4$, and
$\Omega=0.5$. The different lines refer to different initial states: the red
line for $\alpha=1/\sqrt{2}$, the blue line for $\alpha=1/\sqrt{3}$, and the
green line for $\alpha={1}/{2}$. }%
\label{fig3}%
\end{figure}

Adopting the steps developed previously as for a single qubit ($B$)
accelerating in a vacuum cavity, we will treat the other qubit ($A$) as
remaining stationary (in a vacuum cavity). The two vacuum cavities are assumed
to be the same otherwise except that they are separated by a spacelike
distance. The accelerating $B$-qubit is coupled to the scalar field and
evolves according to Eq. (\ref{foe}), the two qubit quantum state thus becomes%
\begin{align}
&  |\Psi_{f}^{(A\overrightarrow{B})}\rangle\nonumber\\
&  =\left(  C_{0}\alpha\left\vert g\right\rangle _{A}\left\vert e\right\rangle
_{B}+C_{1}\beta\left\vert e\right\rangle _{A}\left\vert g\right\rangle
_{B}\right)  \left\vert 0\right\rangle _{A}\left\vert 0\right\rangle
_{B}\nonumber\\
&  -i\left(  C_{1}\alpha\eta_{_{1}}\left\vert g\right\rangle _{A}\left\vert
g\right\rangle _{B}+C_{0}\beta\eta_{_{0}}\left\vert e\right\rangle
_{A}\left\vert e\right\rangle _{B}\right)  \left\vert 0\right\rangle
_{A}\left\vert 1_{k}\right\rangle _{B}.\hskip 18pt \label{1eq}%
\end{align}
Based on the same understanding for the Unruh effect on a single qubit as
studied earlier, the coupling with the scalar field clearly is capable of
causing decoherence to the bipartite entangled state \cite{fm05}, consistent
with the general influence of environment on a quantum state \cite{whz03}. On
the other hand, if the single qubit coherence can be enhanced by the
anti-Unruh effect, one cannot be prevented from speculating that it is also
possible to observe enhanced entanglement between the two qubits for an
accelerating observer accompanying the accelerating $B$ qubit. This is easily
checked. We compute the logarithmic negativity according to Eq. (\ref{lne})
and concurrence given by Eq. (\ref{mui}) for the final bipartite quantum
state, or the reduced density matrix $\rho_{AB}$ obtained by tracing out the
scalar field. The results are presented in Fig. \ref{fig3} for different
initial states. We see that when the interaction timescale $\sigma$ differs
significantly from the timescale associated with the detector gap $\sim
\Omega^{-1}$, enhanced entanglement between the two qubits indeed appears, due
to the anti-Unruh effect. In particular, this result also indicates that the
phenomena of entanglement enhancement is independent of the exact form of the
initial entanglement state, as selected results plotted illustrate different
choices of $\alpha=1/\sqrt{2}$, $1/\sqrt{3}$, and ${1}/{2}$.

One can simply check if the particle exchange symmetry remains true or not
with the accelerated particle labeled by $A$ instead. In contrast to the case
considered above with the particle $B$ being accelerated as in Fig.
\ref{fig3}, in this case, the final two qubit state does become slightly
different although the final conclusion remains the same as before because the
reduced two qubit state observes particle exchange symmetry. For the symmetric
initial state with $\alpha=\beta=1/\sqrt{2}$, everything is symmetric during
all intermediate steps. For other cases presented in Fig. \ref{fig3}, both the
logarithmic negativity and concurrence are employed to check their respective
dependence with the same parameters $\sigma$ and $\Omega$. The results are
consistent with qubit exchange symmetry. In fact, it is easy to check that the
acceleration of either qubit results in the same change to the bipartite
entanglement, since the crucial elements $\eta_{_{0}}$ and $\eta_{_{1}}$,
which determine this change, appear in the same places in the reduced density
matrix $\rho_{AB}$, although the constant parameters $\alpha$ and $\beta$ have
interchanged their roles in the reduced density matrix obtained from tracing
out the scalar field modes from the final state.

\begin{figure}[ptb]
\centering
\includegraphics[width=3.65in]{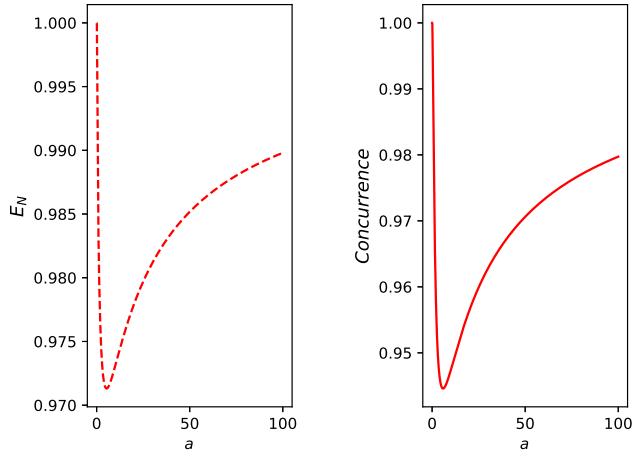} \caption{(Color online) The same as
in Fig. \ref{fig3} for the initial state corresponding to its red line but
with the interaction time prolonged to infinity.}%
\label{fig4}%
\end{figure}

An entangled two qubit state thus cannot hold onto their maximal entanglement
when any one of the two qubits is accelerating. This can be understood from
the view point of entanglement monogamy. If entanglement between two qubits
reaches the maximum, then none of the two qubits can be entangled with any
third party unless the entanglement between the initial two qubits decreases.
When any one qubit accelerates, it will interact with the vacuum. The
entanglement between the accelerated qubit and the vacuum develops, and the
qubit-qubit entanglement decreases. A subsequent question then arises: when
the initial bipartite state is maximal entangled, how could its entanglement
increased further by the anti-Unruh effect? In general, due to the presence of
the switching function, for finite interaction times, even entanglement at
$a=0$ could change, e.g., become non-maximal for the initial maximally
entangled state (see the red line in Fig. \ref{fig3}). This leaves open the
possibility for the discussed enhancement of entanglement. However, the effect
of the switching function disappears when the interaction becomes infinite.
Therefore, for an initial maximally entangled state, it seems that the
anti-Unruh effect cannot enhance its entanglement. The real answer, of course,
is more complicated than this. As seen in Fig. \ref{fig4}, the entanglement
decreases rapidly at the beginning of the acceleration, but it starts to
increase again with the larger acceleration later. Therefore, the anti-Unruh
effect actually still works for infinite interaction time although only for a
finite range of accelerations.

\subsection{Two qubits with the same acceleration}

The previous subsection considers the case of a two qubit system when one of
the qubits is in acceleration. It is found that the two qubit entanglement can
be either increased or decreased, depending on the different conditions. We
now consider for some fixed conditions, e.g. those giving rise to enhanced
entanglement when one of the qubits is in acceleration. What happens to the
qubit-qubit entanglement if the other qubit assumes the same acceleration?
This is an interesting question, since the follow-up observers can remain
stationary respectively with respect to the two qubits simultaneously. For the
Unruh effect, the answer to the above question is a simple resounding
negative, because the acceleration causes only decoherence, as presented in
Ref. \cite{dss15}. But in the presence of the anti-Unruh effect, the situation
becomes subtle and deserves a careful study.

\begin{figure}[ptb]
\centering
\includegraphics[width=3.65in]{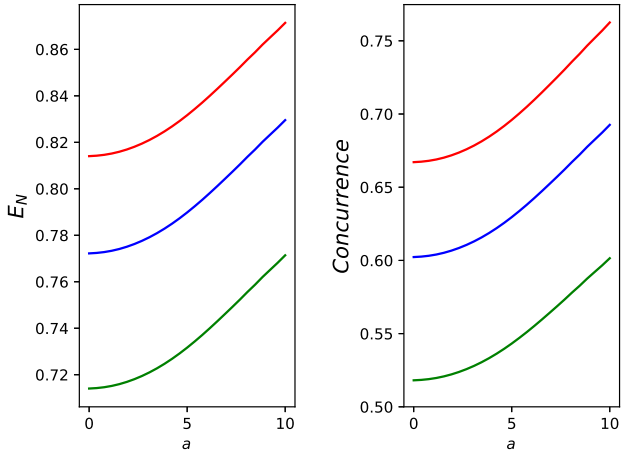} \caption{(Color online) The same as
in Fig. \ref{fig3} but for both qubits in simultaneous acceleration. The
parameters correspond to enhanced entanglement at $\lambda=0.1$, $\sigma=0.4$,
and $\Omega=0.5$. }%
\label{fig5}%
\end{figure}

In this case with the two qubits sharing the same acceleration, the initially
entangled state would evolve into
\begin{align}
&  |\Psi_{f}^{(2)}\rangle\nonumber\\
&  =C_{0}C_{1}[\left(  \alpha\left\vert g\right\rangle _{A}\left\vert
e\right\rangle _{B}+\beta\left\vert e\right\rangle _{A}\left\vert
g\right\rangle _{B}\right)  \left\vert 0\right\rangle _{A}\left\vert
0\right\rangle _{B}\nonumber\\
&  -i\left(  \alpha\eta_{_{1}}\left\vert g\right\rangle _{A}\left\vert
g\right\rangle _{B}+\beta\eta_{_{0}}\left\vert e\right\rangle _{A}\left\vert
e\right\rangle _{B}\right)  \left\vert 0\right\rangle _{A}\left\vert
1_{m}\right\rangle _{B}\nonumber\\
&  -i\left(  \beta\eta_{_{1}}\left\vert g\right\rangle _{A}\left\vert
g\right\rangle _{B}+\alpha\eta_{_{0}}\left\vert e\right\rangle _{A}\left\vert
e\right\rangle _{B}\right)  \left\vert 1_{m}\right\rangle _{A}\left\vert
0\right\rangle _{B}\nonumber\\
&  +\left(  \alpha\eta_{_{0}}\eta_{_{1}}\left\vert e\right\rangle
_{A}\left\vert g\right\rangle _{B}+\beta\eta_{_{0}}\eta_{_{1}}\left\vert
g\right\rangle _{A}\left\vert e\right\rangle _{B}\right)  \left\vert
1_{m}\right\rangle _{A}\left\vert 1_{m}\right\rangle _{B}].\hskip18pt
\end{align}
The logarithmic negativity and concurrence for the reduced density matrix
$\rho_{AB}$ can be analogously computed by first tracing out the scalar field
modes. The results are presented in Fig. \ref{fig5}, which are largely similar
to the results as obtained in Fig. \ref{fig3}, and entanglement enhancement
would occur which is seen to be independent of the initial entangled state
chosen in this case.

Moreover, when the two qubits assume different accelerations, the final
entangled state is found to take a similar form to that for two qubits with
the same acceleration, although the exact values for $\eta_{_{0}}$ and
$\eta_{_{1}}$ depend on these accelerations. The same procedures as employed
in the above can be used to compute the logarithmic negativity and concurrence
for the reduced density matrix, and analogous figures can be obtained by
fixing the acceleration for one qubit while varying the acceleration for the
other qubit. Since nothing particularly new arises for this case, the detailed
results will not be presented here.

\section{Two qubit product state}

Our discussions for the two qubit example in the above section show that
qubit-qubit entanglement can be increased when the anti-Unruh effect is taken
into consideration. This section intends to show that the mechanism behind
enhanced entanglement is more like an entanglement amplifier \cite{cdl01} when
the two qubits share the same acceleration. It is not an entanglement
generator since the initial two qubit entanglement must be nonzero. This
prompts us to study the simple question if one can make two qubits, which
possess no correlation initially and are separated far away from each other,
to become correlated when one or both of the two qubits accelerate? For this
example, we take the initial state with $\alpha=1$ and $\beta=0$ from Eq.
(\ref{eqs}) and first consider the situation when the $B$ qubit is in
acceleration. Like the case considered earlier, the state then becomes%
\begin{align}
&  |\Psi_{f}^{(B)^{\prime}}\rangle\nonumber\\
&  =C_{1}\left(  \left\vert g\right\rangle _{A}\left\vert e\right\rangle
_{B}\left\vert 0\right\rangle _{A}\left\vert 0\right\rangle _{B}-i\eta_{_{1}%
}\left\vert g\right\rangle _{A}\left\vert g\right\rangle _{B}\left\vert
0\right\rangle _{A}\left\vert 1_{k}\right\rangle _{B}\right)  .\nonumber\\
&
\end{align}
Tracing out the scalar field modes, we obtain the final state for the two
qubits
\begin{align}
\rho_{AB}^{\prime}  &  =\left\vert g\right\rangle _{A}\left\langle
g\right\vert _{A}\otimes\left\vert C_{1}\right\vert ^{2}\left(  \left\vert
e\right\rangle _{B}\left\langle e\right\vert _{B}+\eta_{_{1}}^{2}\left\vert
g\right\rangle _{B}\left\langle g\right\vert _{B}\right) \nonumber\\
&  =\rho_{A}\otimes\rho_{B},
\end{align}
obviously this is still a product state. The same result arises if qubit $A$
is assumed to be in acceleration instead. Thus entanglement creation from
accelerating one qubit is impossible.

What happens when both qubits are accelerated simultaneously? For simplicity,
we still consider the same acceleration for the two qubits, and the initial
product state $\left\vert \Psi_{i}^{\prime}\right\rangle =\left\vert
g\right\rangle _{A}\left\vert e\right\rangle _{B}\left\vert 0\right\rangle
_{A}\left\vert 0\right\rangle _{B}$ then evolves into%
\begin{align}
&  |\Psi_{f}^{\prime}\rangle\nonumber\\
&  =C_{0}C_{1}(\left\vert g\right\rangle _{A}\left\vert e\right\rangle
_{B}\left\vert 0\right\rangle _{A}\left\vert 0\right\rangle _{B}-i\eta_{_{1}%
}\left\vert g\right\rangle _{A}\left\vert g\right\rangle _{B}\left\vert
0\right\rangle _{A}\left\vert 1_{m}\right\rangle _{B}\nonumber\\
&  -i\eta_{_{0}}\left\vert e\right\rangle _{A}\left\vert e\right\rangle
_{B}\left\vert 1_{m}\right\rangle _{A}\left\vert 0\right\rangle _{B}%
-\eta_{_{0}}\eta_{_{1}}\left\vert e\right\rangle _{A}\left\vert g\right\rangle
_{B}\left\vert 1_{m}\right\rangle _{A}\left\vert 1_{m}\right\rangle
_{B}).\nonumber\\
\end{align}
Again, the reduced two qubit density matrix is obtained by tracing out the
scalar field modes and we obtain
\begin{align}
\rho_{AB}^{\prime}  &  =\rho_{A}\otimes\rho_{B},
\end{align}
i.e., no entanglement is created.

Based on the perturbation treatment for the two qubit evolution dynamics with
accelerations, we thus find entanglement cannot be created for an initial
product state, but can be enhanced for any initial state with entanglement
through the anti-Unruh effect. We believe that the main reason for this
conclusion lies at our treatment of the vacuum as a product state. If the two
qubits are also in a product state, they cannot influence each other when one
or both qubits interact with the vacuum through accelerated motion since no
channel can be used to transfer the influence. On the other hand, however, as
discussed before \cite{rrs05,sm09,mm11}, if the vacuum state is entangled, the
initial product state for the two qubits can become entangled since the vacuum
can provide the channel to facilitate the interchange of their respective
information and coherence. When the initial two qubit state is entangled, even
if the vacuum state is not, the two qubits can still interchange their
respective change in the coherence through the channel facilitated by their
initial entanglement. When the two qubits dynamics are the same, the change of
their respective coherence will also be the same. Thus, the change of the two
qubit entanglement is essentially the same as the change of the single qubit
coherence, both can be derived from the interaction between the accelerated
qubit(s) and the vacuum. This shows that the enhancement of the entanglement
is indeed caused by the anti-Unruh effect recently discussed.

\section{Conclusion}

In this paper, we revisit the anti-Unruh effect for single qubit state from
the change of the decoherence factor due to acceleration. We extend such a
discussion to two qubits state, and show that the enhanced entanglement
between the two qubits can occur aided by the anti-Unruh effect, not only for
the case when one of the qubits is in acceleration, but also for the case when
both qubits are accelerating simultaneously. To avoid any possible
misunderstanding that enhanced enhancement is derived from the exchange with
the vacuum entanglement, we always model the vacuum in our calculation as in a
product state. Thus, when the initial bipartite state is also taken as a
product state, the acceleration of the qubit(s) cannot create any entanglement
between the two qubits, even though the anti-Unruh effect remains present.
This is reasonable, since the local effect caused by acceleration also
requires the channel to transfer the influence to the place where the other
qubit is located, no matter whether this channel is formed by the entanglement
of the bipartite state itself or through the vacuum entanglement. Thus, we
conclude that the anti-Unruh effect can only lead to entanglement enhancement
for an initially entangled state, but cannot create entanglement out of a
non-entangled state.

\section{Acknowledgement}

This work is supported by the NSFC grant No. 11654001. B.C.Z also acknowledges
the NSFC support (grant No. 11374330 and No. 91636213). The work of L.Y. is
also supported by the MOST 2013CB922004 of the National Key Basic Research
Program of China and by NSFC (grant No. 11374176).

\end{document}